\documentclass[journal]{IEEEtran}
\usepackage{amsmath,amsfonts,amssymb}
\usepackage{graphicx}
\usepackage{cite}
\usepackage{mathptmx}
\usepackage{multirow}
\usepackage{multicol}

\begin{document}

\title{Anti-Forensics of Camera Identification and the Triangle Test by Improved Fingerprint-Copy Attack}

\author{
Haodong~Li,
Weiqi~Luo,
Quanquan~Rao,
Jiwu~Huang
\thanks{H. Li is with the School of Electronics and Information Technology, Sun Yat-sen University, Guangzhou, China.}
\thanks{W. Luo (corresponding author) is with the School of Data and Computer Science, and Guangdong Key Laboratory of Information Security Technology, Sun Yat-sen University, Guangzhou, China (e-mail: luoweiqi@mail.sysu.edu.cn).}
\thanks{Q. Rao is with the School of Data and Computer Science, Sun Yat-sen University, Guangzhou, China.}
\thanks{J. Huang is with the College of Information Engineering, Shenzhen University, Shenzhen, China.}
}

\maketitle

\begin{abstract}
The fingerprint-copy attack aims to confuse camera identification based on sensor pattern noise. However, the triangle test shows that the forged images undergone fingerprint-copy attack would share a non-PRNU (Photo-response nonuniformity) component with every stolen image, and thus can detect fingerprint-copy attack. In this paper, we propose an improved fingerprint-copy attack scheme. Our main idea is to superimpose the estimated fingerprint into the target image dispersedly, via employing a block-wise method and using the stolen images randomly and partly. We also develop a practical method to determine the strength of the superimposed fingerprint based on objective image quality. In such a way, the impact of non-PRNU component on the triangle test is reduced, and our improved fingerprint-copy attack is difficultly detected. The experiments evaluated on 2,900 images from 4 cameras show that our scheme can effectively fool camera identification, and significantly degrade the performance of the triangle test simultaneously.
\end{abstract}


\section{Introduction}
\label{sec:Intro}

With the various powerful image editing software, such as GIMP and Adobe Photoshop, modifying digital images becomes increasingly easy, which would inevitably lead to some potential moral and/or legal consequences. Nowadays, digital image forensics become an important issue \cite{Stamm2013}. Image forensics involve many technical methods for different applications such as detecting image compression \cite{Farid2009,Lin2009,Luo2010}, exposing image processing history \cite{Popescu2005,Yuan2011,Li2013}, revealing tampered images \cite{He2012,Qiu2014,LiangYDL2015}, differentiating between photorealistic and photographic images \cite{Lyu2005,Dirik2007}, identifying the acquisition component or device of an image \cite{Lukavs2006a,Khanna2007,Swaminathan2007,Celiktutan2008}, and so on. As one of the important issues in image forensics, source camera identification aims to identify which camera was used to capture the given image. Kurosawa \emph{et al.} \cite{Kurosawa1999} carried out the identification utilizing fixed pattern noise (FPN) appearing on dark frames; Kharrazi \emph{et al.} \cite{Kharrazi2004} proposed a feature set based on the color of pixel and image quality metrics. In \cite{Lukavs2006a}, Luk\'{a}\v{s} \emph{et al.} proposed a novel source camera identification method based on PRNU (Photo-response nonuniformity). By averaging the noise components of sufficient images from a given camera, the obtained result can be regarded as an estimation of the PRNU component, which can serve as a reliable and unique fingerprint for the camera. In \cite{Chen2008}, Chen \emph{et al.} proposed a method to estimate the PRNU factor based on maximum-likelihood approach, and shown that the method outperformed that proposed in \cite{Lukavs2006a}. Besides, several improved PRNU-based methods have been proposed for source camera identification. For example, Li \cite{Li2010} proposed an enhanced method to suppress some unwanted components within the noise components. Hu \emph{et al.} \cite{Hu2010} proposed a way to construct camera fingerprints from all of the three color channels via using the characteristics of the color filter array (CFA) structure. Kang \emph{et al.} \cite{Kang2012} introduced a new detection method to lower the false positive rate. Bayram \emph{et al.} \cite{Bayram2012} tried to reduce the computational expensive of conventional approaches through binarizing the sensor fingerprints. In \cite{F.Gharibi2010} and \cite{Li2014}, two methods are proposed to estimate the sensor fingerprints based on local information of image via block-wise strategies.

On the opposite side of forensics, however, a wise attacker can perform some anti-forensic operations to confuse the corresponding forensic methods. Studying the anti-forensic methods is very important for researchers, since it can help to discover the limitations of current forensic methods and further improve their performance. Up to now, several typical anti-forensic works have been reported. For instance, Stamm and Liu \cite{Stamm2011} proposed a scheme to conceal the traces left by digital image compression. In \cite{Gloe2007} and ~\cite{Boehme2013}, some methods are proposed to trick the resampling detector \cite{Popescu2005} and the sensor pattern noise based image forgeries detector \cite{Lukavs2006a}, respectively. In \cite{Dirik2014}, the authors proposed a method to remove the PRNU noise within an image so that it would not be matched with its source camera. Some methods such as \cite{Goljan2011} and \cite{Caldelli2011} try to confuse the source camera identification as described above via \emph{fingerprint-copy attack}. These methods firstly estimate the fingerprint of a camera $\mathcal{C}_A$ from some stolen images, and then superimpose it into a target image taken by a different camera $\mathcal{C}_E$ to disguise the resulting image as one taken by camera $\mathcal{C}_A$. Most existing PRNU based camera identification algorithms would incorrectly identify that the resulting forged image is taken by camera $\mathcal{C}_A$ rather than camera $\mathcal{C}_E$.

Forensics and anti-forensics is like the cat-and-mouse game. Recently, Goljan \emph{et al.} \cite{Goljan2011} proposed the \emph{triangle test} method to expose the fingerprint-copy attack. They pointed out that there would be a common non-PRNU component shared by those stolen images and the resulting forged images after performing the fingerprint-copy attack. Based on such a property, it can identify the stolen images from some candidate images using the individual test, determine whether a set of candidate images contain some stolen images using the pooled test, and directly detect the forged images using the multiple forgeries test.

In this paper, we propose an anti-forensic method to deceive both camera identification and the triangle test. The main idea is to randomly select just a small part of stolen images for fingerprint estimation, and then superimpose the estimated fingerprint into the target image block by block. By doing so, we are able to reduce impact of the non-PRNU components from the stolen images, and thus it would deceive the triangle test. Moreover, we propose a practical way to determine the strength of the superimposed fingerprint based on adjusting PSNR, so we can overcome the drawback of conventional fingerprint-copy attack that the true fingerprint needs to be available. In our experiments, 2,900 images from three different camera brands (four camera individuals) and three different kinds of the triangle tests have been included. The experimental results have shown that the proposed method outperforms the conventional fingerprint-copy attack method \cite{Goljan2011} and degrades the performance of triangle test significantly.

The rest of this paper is organized as follows. Section \ref{Sec:Triangle Test} gives a brief overview of some related techniques on camera identification based on PRNU. Section \ref{Sec:Proposed_Method} presents the proposed anti-forensic method, including superimpose the estimated fingerprint dispersedly and determine the fingerprint strength according to PSNR. Section \ref{Sec:Experimental} shows the experimental results and discussions. Finally, the concluding remarks and future works are given in Section \ref{Sec:Conclusion}.

\section{Related Works}
\label{Sec:Triangle Test}

In this section, we will introduce some related works including the camera identification method based on PRNU, the conventional fingerprint-copy attack method and the triangle test method. Beforehand, some notations used in this paper are given: everywhere a boldface symbol represents a matrix, \emph{e.g.} ${\bf X}$; for two matrices ${\bf X}$ and ${\bf Y}$ with the same dimensions, their element-wise product and  element-wise division are denoted as ${\bf Z}={\bf X}{\bf Y}$ and ${\bf Z}={\bf X}/{\bf Y}$ respectively.

\subsection{Camera Identification Based on PRNU}
\label{subsec:threshold_based}
Due to the manufacturing imperfection and the nonuniformity of silicon wafers, the recorded pixels on camera sensors vary from each other even if the sensors are exposed to the same illumination. Therefore, a kind of pattern noise called PRNU exists in all sensor-based cameras, and it is introduced to every image taken by camera. Such noise is independent of the environment. More importantly, the noises of all images taken by a same camera are highly correlated, while the correlations among the noises of those images taken by different cameras would be much weaker. Based on this property, therefore, the pattern noise can be regarded as a unique fingerprint of a digital camera.

In \cite{Chen2008}, Chen \emph{et al.} proposed a maximum likelihood method to estimate the PRNU multiplicative factor $\hat{\bf K}$ of a camera $\mathcal{C}_A$
\begin{equation}\label{eqt:estimate}
  \hat{\bf K} = \frac{\sum\limits_{i = 1}^N {{{\bf{W}}_i}{{\bf{I}}_i}}}{\sum\limits_{i = 1}^N {{\bf I}_i^2}}
\end{equation}
where ${\bf{I}}_1,\dots,{\bf I}_N$ are the $N$ images taken by camera $\mathcal{C}_A$ that are used to estimate the fingerprint, and {${\bf{W}}_i = {\bf{I}}_i- F({\bf{I}}_i)$} is the noise residual of ${\bf{I}}_i$, where $F$ is the wavelet denoising filter \cite{Mihcak1999}. For a given image $\bf J$, the presence of the camera fingerprint can be evaluated by the correlation detector:
\begin{equation}\label{eqt:corr}
  \rho = corr({\bf W_J},{\bf J}\hat{\bf K})
\end{equation}
where $\hat{\bf K}$ is the estimated fingerprint of camera $\mathcal{C}_A$ computed by Eq. (\ref{eqt:estimate}). Thus, camera identification can be achieved with a detector based on the statistic $\rho$. By setting a false alarm rate $P_{fa}$ (\emph{i.e.} the probability of that those images taken by other cameras are identified as being taken by camera $\mathcal{C}_A$), some training images taken by other cameras are used to determine the detection threshold $t_1$. Then for a testing image $\bf J$, its corresponding $\rho$ is obtained based on Eq. (\ref{eqt:corr}). If $\rho > t_1$, $\bf J$ is decided to be taken by camera $\mathcal{C}_A$; otherwise, $\bf J$ is not taken by camera $\mathcal{C}_A$. It is noted that the PRNU based correlation detection is an important method in camera identification, its reliability and effectiveness have been verified in \cite{Lukavs2006a} and \cite{Chen2008}.

\subsection{Fingerprint-copy Attack}
\label{subsec:Fingerprint-copy-Attack}
The purpose of fingerprint-copy attack \cite{Goljan2011} is to confuse those threshold-based camera identification algorithms. Assume that Alice is the victim and Eve is the attacker. Alice has posted some images taken by her camera $\mathcal{C}_A$ on some websites, and Eve steals $N$ of these images. In order to make a forgery, Eve can firstly suppress the fingerprint within a target image $\rm\bf J$ taken by a different camera $\mathcal{C}_E$. Next, she estimates a fake fingerprint $\hat{\bf K}_{\rm E}$ of camera $\mathcal{C}_A$ with Eq. (\ref{eqt:estimate}) using the stolen images. Finally, she superimposes the fake fingerprint $\hat{\bf K}_{\rm E}$ into $\rm\bf J$ and makes a forgery ${\bf J}^\prime$ as
\begin{equation}\label{eqt:fingerprint_copy}
  {\bf J}^\prime = [{\bf J}(1 + \alpha \hat{\bf K}_{\rm E})]
\end{equation}
where the symbol $[\cdot]$ is the rounding and truncating operation (a real number is rounded to its nearest integer, and it will be truncated to 0 or 255 if it is outside the range of [0,255]), and the parameter $\alpha>0$ denotes the fingerprint strength. In order to create a good forgery, Eve must set a proper parameter $\alpha$ to make the response of the camera identification detector (\emph{e.g.} the correlation detector using Eq. (\ref{eqt:corr})) on ${\bf J}^\prime$ is ``natural'' as if ${\bf J}^\prime$ was indeed taken by camera $\mathcal{C}_A$. In this case, the fingerprint strength $\alpha$ should be adjusted according to the true fingerprint $\bf K$ of camera $\mathcal{C}_A$.

Please note that two issues should be further considered in this fingerprint-copy attack strategy. Firstly, is it a good strategy to use all the $N$ stolen images for estimating the whole fake fingerprint $\hat{\bf K}_{\rm E}$?  Secondly, how to adjust the fingerprint strength $\alpha$ to make the correlation computed by Eq. (\ref{eqt:corr}) on ${\bf J}^\prime$ ``natural''? Since Eve cannot access the true fingerprint $\bf K$ of camera $\mathcal{C}_A$, it seems unreasonable for her to obtain a natural $\alpha$ in real applications. To avoid this assumption, we will develop a method in Section \ref{Sec:Proposed_Method} via adjusting PSNR to determine the proper fingerprint strength.

\subsection{Triangle Test}
\label{subsec:triangle}
In order to prove her innocence, Alice can perform the triangle test \cite{Goljan2011} to expose the existence of fingerprint-copy attack.

\begin{figure}[t]\centering
{\includegraphics[scale=1.5]{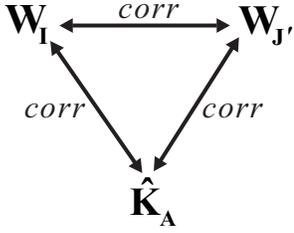}}
\caption{Illustration of the triangle test.}
\label{fig:Triangle}
\end{figure}

One possible scenario is that Alice only gets a forged image ${\bf J}^\prime$, and all the $N$ images stolen by Eve belong to a large database, containing $N_c~(N_c > N)$ images, which is available to Alice. In such a case, Alice tries to identify the stolen images from the image database so as to prove ${\bf J}^\prime$ is a forgery.
Since Alice has the camera $\mathcal{C}_A$, she can obtain a good estimated fingerprint $\hat{\bf K}_{\rm A}$ of camera $\mathcal{C}_A$ via Eq. (\ref{eqt:estimate}) using some well designed images (\emph{e.g.} blue sky images) which are not in the image database $N_c$. Then for each candidate image $\bf I$ in the database, she computes the three correlations between the noise residual of ${\bf I}$, the noise residual of ${\bf J}^\prime$, and the estimated fingerprint $\hat{\bf K}_{\rm A}$, denoted as $c_{{\bf I},{\bf J}^\prime}$, $c_{{\bf I},\hat{\bf K}_{\rm A}}$ , and $c_{{\bf J}^\prime,\hat{\bf K}_{\rm A}}$ respectively, as shown in Fig.~\ref{fig:Triangle}.

In \cite{Goljan2011}, the authors obtained an estimation of  $c_{{\bf I},{\bf J}^\prime}$ (denoted as $\hat c_{{\bf I},{\bf J}^\prime}$) from the two other correlations  $c_{{\bf I},\hat{\bf K}_{\rm A}}$ and $c_{{\bf J}^\prime,\hat{\bf K}_{\rm A}}$. They further showed that for a forged image ${\bf J}^\prime$, the $c_{{\bf I},{\bf J}^\prime}$ and $\hat{c}_{{\bf I},{\bf J}^\prime}$ were highly linear dependent if $\bf I$ was not used by Eve. If the image $\bf I$ was previously used by Eve, however, such a linear property would be destroyed and $c_{{\bf I},{\bf J}^\prime}$ would present a higher value (Please refer to Fig.~\ref{figure:triangle test 1}(a) for example), since $\bf I$ and ${\bf J}^\prime$ share a non-PRNU component in this case. Therefore, via fitting a straight line $c_{{\bf I},{\bf J}^\prime}=\lambda \hat c_{{\bf I},{\bf J}^\prime}+\eta$ ($\lambda$ and $\eta$ are the fitting parameters) using some images that have not been used by Eve and obtaining the pdf (Probability Density Function): $f_{{\bf J}^\prime}(x) \approx {\rm Pr}(c_{{\bf I},{\bf J}^\prime}-\lambda \hat c_{{\bf I},{\bf J}^\prime}-\eta = x|\hat c_{{\bf I},{\bf J}^\prime})$, Alice has two alternative options to perform the triangle test as follows.

\begin{enumerate}
\item[(a)] The individual test. Alice can set a threshold $t_2$ on a certain false alarm rate, and test each candidate image $\bf I$ individually by evaluating $c_{{\bf I},{\bf J}^\prime}-\lambda \hat c_{{\bf I},{\bf J}^\prime}-\eta$. If the value is larger than $t_2$, then $\bf I$ is used by Eve.
\item[(b)] The pooled test, namely, testing all $N_c$ candidate images at once whether $c_{{\bf I},{\bf J}^\prime}-\lambda \hat c_{{\bf I},{\bf J}^\prime}-\eta \sim f_{{\bf J}^\prime}(x)$ is satisfied. If the answer is not, there would be at least one stolen image among those candidate ones.
\end{enumerate}

Another possible scenario is that Alice may access more than one forged image such as ${\bf J}_1^\prime$ and ${\bf J}_2^\prime$, and it is assumed that these images are created with a same fake fingerprint $\hat{\bf K}_{\rm E}$. In this case, since ${\bf J}_1^\prime$ and ${\bf J}_2^\prime$ would share a non-PRNU component coming from $\hat{\bf K}_{\rm E}$ and some common noises introduced by the camera $\mathcal{C}_E$, Alice can reveal the forgeries by running the triangle test on ${\bf W}_{{\bf J}_1^\prime}$, ${\bf W}_{{\bf J}_2^\prime}$, and $\hat{\bf K}_{\rm A}$ without accessing the $N$ stolen images used by Eve.

\section{the Proposed Fingerprint-copy Attack Scheme}
\label{Sec:Proposed_Method}

To design a successful fingerprint-copy attack, the two following requirements should be carefully considered.

\begin{enumerate}
\item[(a)] Confuse the camera identification. Since Eve's original intention is to frame Alice as the owner of the forged image ${\bf J}^\prime$. Therefore, after superimposing a fake fingerprint into a target image, it should guarantee that the camera identification detectors cannot distinguish the forged image ${\bf J}^\prime$ from the original ones.
\item[(b)] Fool the triangle test. To avoid being detected, the performance of the triangle test evaluated on the resulting forgeries should degrade significantly after using the fingerprint-copy attack.
\end{enumerate}

There is a tradeoff between the two requirements. Usually, the more strength of a fake fingerprint is superimposed, the easier to confuse the camera identification (\textit{i.e.} requirement \#a), however, the better detection performance of the triangle test (\textit{i.e.} requirement \#b) would be obtained. Based on the analysis in Section \ref{subsec:triangle}, we find that the key idea of the triangle test is to detect the non-PRNU component from the stolen images. If we can reduce the impact of the non-PRNU component on the triangle test, it is expected that the performance of the triangle test will drop significantly.

\subsection{A Block-wise and Randomized Fingerprint-copy Scheme}
\label{subsec:weaken}
Unlike the conventional fingerprint-copy attack described Section \ref{subsec:Fingerprint-copy-Attack}, in this subsection we introduce a block-wise and randomized fingerprint-copy scheme. Assuming that we have stolen $N$ images taken by Alice's camera, we produce the fake fingerprint for different blocks of the target image by just using a randomly selected subset of the $N$ stolen images. The details are as follows.

Firstly, we divide all stolen images ${\bf I}_1, \ldots, {\bf I}_N$ (assuming that their sizes are the same as the target one) and the target image $\bf J$ into $n$ non-overlapping blocks with the size of $l\times l$. For each block ${\bf J}_{b}$ of image $\bf J$, where $b=1,2,\ldots,n$, we randomly select $r~(r<N)$ stolen images ${{\bf{I}}^\prime}_{1}, \ldots, {{\bf{I}}^\prime}_{r}$ and calculate $\hat{\bf{K}}_{{\rm E},b}$ according to Eq. (\ref{eqt:estimate}) using the corresponding blocks ${{\bf{I}}^\prime}_{1,b} \ldots  {{\bf{I}}^\prime}_{r,b}$ of the selected stolen images, then we superimpose this fake fingerprint block $\hat{\bf{K}}_{{\rm E},b}$ into the corresponding block ${\bf J}_{b}$ within the target image,
\begin{equation}\label{eqt:fingerprint_copy_block}
  {\bf J}_b^\prime = [{{\bf J}_b^{}}(1 + \alpha_b \hat{\bf K}_{{\rm E},b})]
\end{equation}
where $\alpha_b$ is the fingerprint strength for the $b$th block. We will discuss how to select a proper $\alpha_b$ in Section \ref{subsec:setalpha}.

\begin{figure*}[t]
  \centering
  \includegraphics[width=4.5cm]{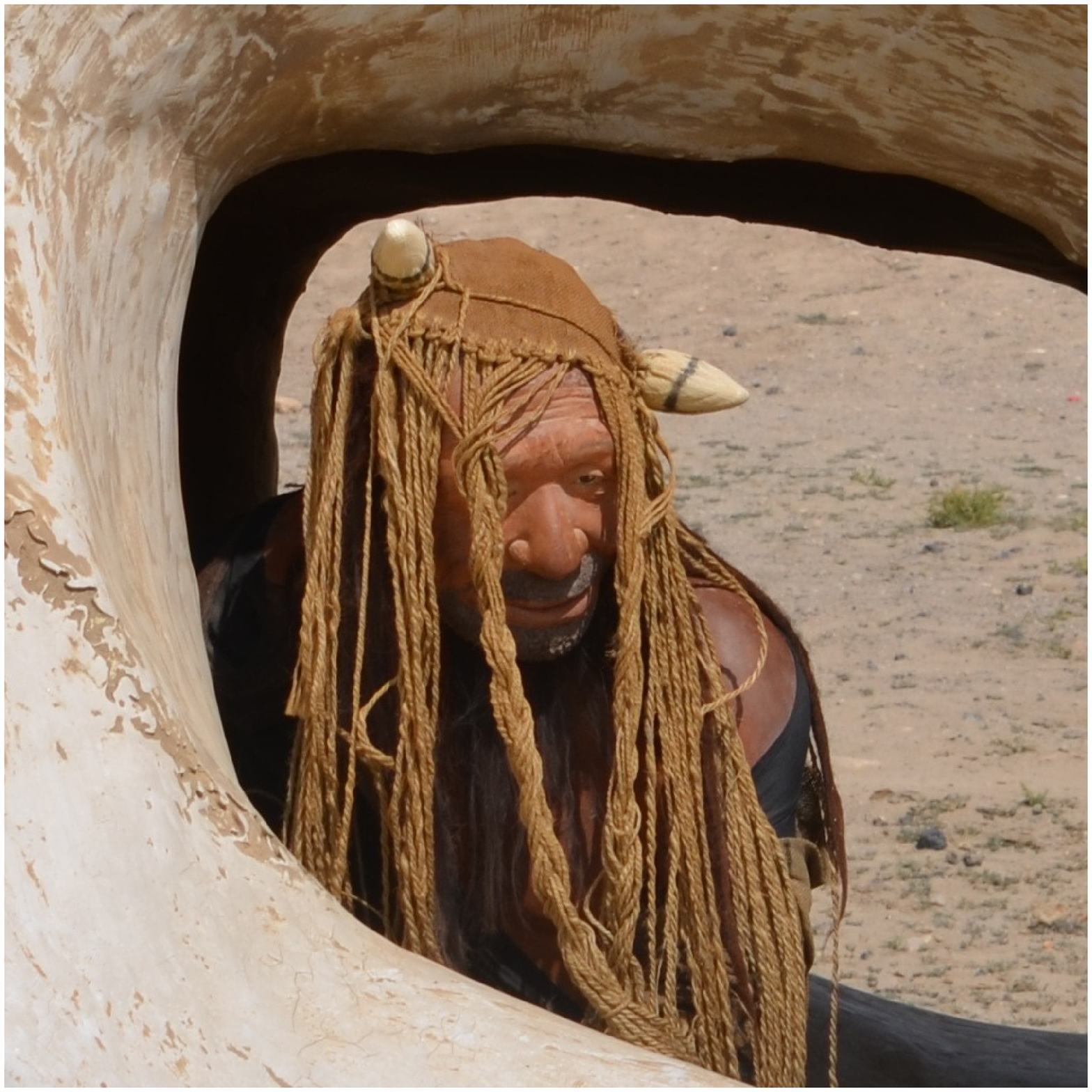}\hspace{1em}
  \includegraphics[width=4.5cm]{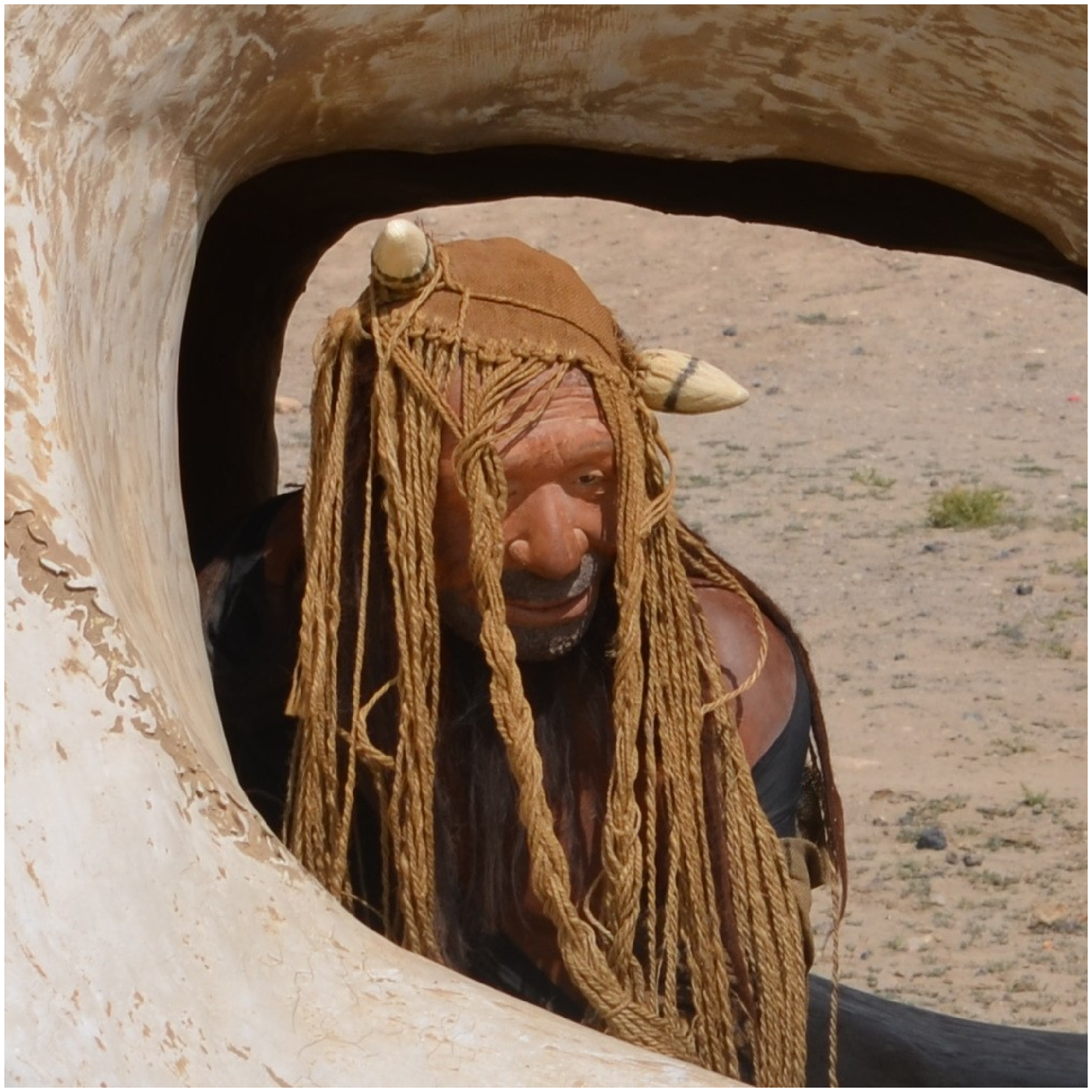}
  \caption{Original image (left) taken by a Nikon D7000 camera and the forged image (right) created by the proposed method with $N=100$, $A=50$ and $r=10$.}
  \label{fig:forged example}
\end{figure*}

In such a way, the fake fingerprint superimposed into the target image $\bf{J}$ is dispersed, so the common non-PRNU component between each stolen image ${\bf I}_i$ and the forged image $\bf{J^\prime}$ is also dispersed. As a result, it is expected that the impact of the non-PRNU component on the triangle test will be reduced. Based on our experiments (please see Section \ref{subsec:Individual} to Section \ref{subsec:mutiple test}), the performance of the triangle test is significantly degraded when applying the proposed fingerprint-copy scheme. On the other hand, it is noted that the dispersion of fingerprint has slight effect on the performance for confusing the camera identification. Based on our experimental results (please see Section \ref{subsec:confusion}), at the same level of PSNR, the proposed method still achieves similar results for camera identification compared to the conventional method \cite{Goljan2011}.


\subsection{Determining the Fingerprint Strength}
\label{subsec:setalpha}
An important factor in Eq. \ref{eqt:fingerprint_copy_block} is the fingerprint strength $\alpha_b$. Since a forger does not know the true fingerprint $\bf{K}$ in practice, we try to  determine $\alpha_b$ based on the objective image quality measured by PSNR as the following equation,

\begin{equation}\label{eqt:alpha_b}
\alpha _b^* \in \{\alpha_b|\textmd{PSNR}({\bf J}_b^\prime,{\bf J}_b^{})= A\}
\end{equation}
where $\textmd{PSNR}({\bf J}_b^\prime,{\bf J}_b^{})$  stands for the PSNR between an image block after and before superimposing the fake fingerprint, and $A$ is a target PSNR value. In other words, we select $\alpha _b^*$ so that the PSNR between the forged image block ${\bf J}_b^\prime$ and the source image block ${\bf J}_b^{}$ is equal to the given PSNR value $A$. In practice, it is not easy to obtain the solution $\alpha _b^*$ directly. Besides, there would be several $\alpha _b$ that satisfy $\textmd{PSNR}({\bf J}_b^\prime,{\bf J}_b^{})= A$  due to the rounding/truncation errors incurred in generating ${\bf J}_b^\prime$. Therefore, we use binary search to approximately find out $\alpha _b^*$, which can ensure $\textmd{PSNR}({\bf J}_b^\prime,{\bf J}_b^{})$ is as close to $A$ as possible.

Based on the work in \cite{Goljan2011}, we note that the PSNR between the original image and the forged image after the fingerprint-copy attack with the ``natural'' parameter $\alpha$  would approximately fall in the range of [47.6 dB, 58.7 dB]. Therefore, we can select the value of $A$ within this range. Usually, the larger the $A$ to be selected, the smaller the $\alpha_b^*$ is obtained, and vice versa.

\subsection{Implementation of the Proposed Method}
\label{subsec:proposed_method}

Based on the experiments and discussions in the Section \ref{subsec:weaken} and \ref{subsec:setalpha}, we summarize the procedures for creating a forged image ${\bf J}^\prime$ as follows.

\begin{itemize}
  \item [(a)] Divide the target image $\bf J$ and all the stolen images ${\bf I}_1, \ldots, {\bf I}_N$ into $n$ non-overlapping small blocks. For each block ${\bf J}_b^{}~(b=1,2,\ldots,n)$, randomly select $r~(r\leq N)$ corresponding stolen image blocks ${{\bf{I}}^\prime}_{1,b}, \ldots, {{\bf{I}}^\prime}_{r,b}$ to estimate $\hat{\bf{K}}_{{\rm E},b}$ according to Eq. (\ref{eqt:estimate}).

  \item [(b)] For each block ${\bf J}_b^{}$, we set $A$ (the target PSNR) for adjusting fingerprint strength, and then we find the proper fingerprint strength $\alpha_b^*$ according to Eq. (\ref{eqt:alpha_b}), finally superimpose the estimated fingerprint $\hat{\bf{K}}_{{\rm E},b}$ into ${\bf J}_b^{}$  with Eq. (\ref{eqt:fingerprint_copy_block}) (set $\alpha_b$ as $\alpha_b^*$) to obtain a forged block ${\bf J}_b^\prime$.

  \item [(c)] After processing all the blocks ${\bf J}_b^{}~(b=1,2,\ldots,n)$ as described in step (b), we combine all ${\bf J}_b^\prime~(b=1,2,\ldots,n)$ together to generate a forged image ${\bf J}^\prime$.
\end{itemize}

By choosing proper parameters (\emph{i.e.} $r$ and $A$), we can achieve a good tradeoff between the two requirements described previously.
Comparing with the conventional fingerprint-copy attack \cite{Goljan2011}, the main difference of the proposed method described above is that we superimpose the estimated fingerprint into a target image block by block rather than the whole image, and we just randomly use some stolen images to estimate the corresponding fingerprint for each forged image block rather than all stolen images. Furthermore, we obtain the proper fingerprint strength via adjusting the PSNR rather than using a natural $\alpha$ which is not available in practice.

An image forgery made by the proposed method with parameter $N=100$, $A=50$  and $r=10$ is illustrated in Fig.~\ref{fig:forged example}. Since the PSNR between this forgery and its original counterpart is 50 dB, no perceptible artifacts are introduced in the resulting image. This forgery can successfully confuse the camera identification and fool the triangle test simultaneously. More comparative results will be given in the Section \ref{Sec:Experimental}.

\section{Experimental Results}
\label{Sec:Experimental}

All our experiments are conducted on 2,900 images taken by four different digital cameras, including Nikon D7000 \#1, Nikon D7000 \#2, Canon EOS 400D, and Pentax K20D. The number of images for each camera and their imaginary owners are shown in Table~\ref{table:dataset}. Please note that there are two digital cameras with the same model, Nikon D7000 (\textit{i.e.} $\mathcal{C}_A$ and $\mathcal{C}_{E1}$), since Eve may create the image forgery according to the model of Alice's camera to avoid introducing other detectable artifacts, such as inconsistent CFA \cite{Popescu2005a} and quantization matrix \cite{Farid2006}. Besides, camera $\mathcal{C}_{E2}$ and $\mathcal{C}_{E3}$ with different brands and models are also included in our experiments.

\begin{table*}[t]
\centering
\caption{Image databases used in our experiments}{
\begin{tabular}{c|c|c|c} \hline
  Camera name  & Camera brand & Number of images & Owner\\ \hline
  $\mathcal{C}_A$  & Nikon D7000 \#1 & 2000 & Alice\\ \hline
  $\mathcal{C}_{E1}$ & Nikon D7000 \#2 & 300  & Eve  \\ \hline
  $\mathcal{C}_{E2}$ & Canon EOS 400D  & 300  & Eve  \\ \hline
  $\mathcal{C}_{E3}$ & Pentax K20D     & 300  & Eve  \\ \hline
\end{tabular}}
\label{table:dataset}
\end{table*}

\begin{table*}[t]
\centering
\caption{$P_{fa}$ (\%) and $P_D$ (\%) for different block size}{
\begin{tabular}{c|ccccccc} \hline
  $l$ & 8 & 16 & 32 & 64 & 128 & 256 & 512 \\ \hline
  $P_{fa}$ & 96.33 & 99.33 & 100 & 100 & 100 & 100 & 100 \\ \hline
  $P_D$ & 24.56 & 44.22 & 54.19 & 58.29 & 56.55 & 49.35 & 31.21 \\ \hline
\end{tabular}}
\label{table:select_l}
\end{table*}

\begin{figure*}[t]
  \centering
  \begin{minipage}[c]{0.48\textwidth}
  \centering
  \includegraphics[width=7cm]{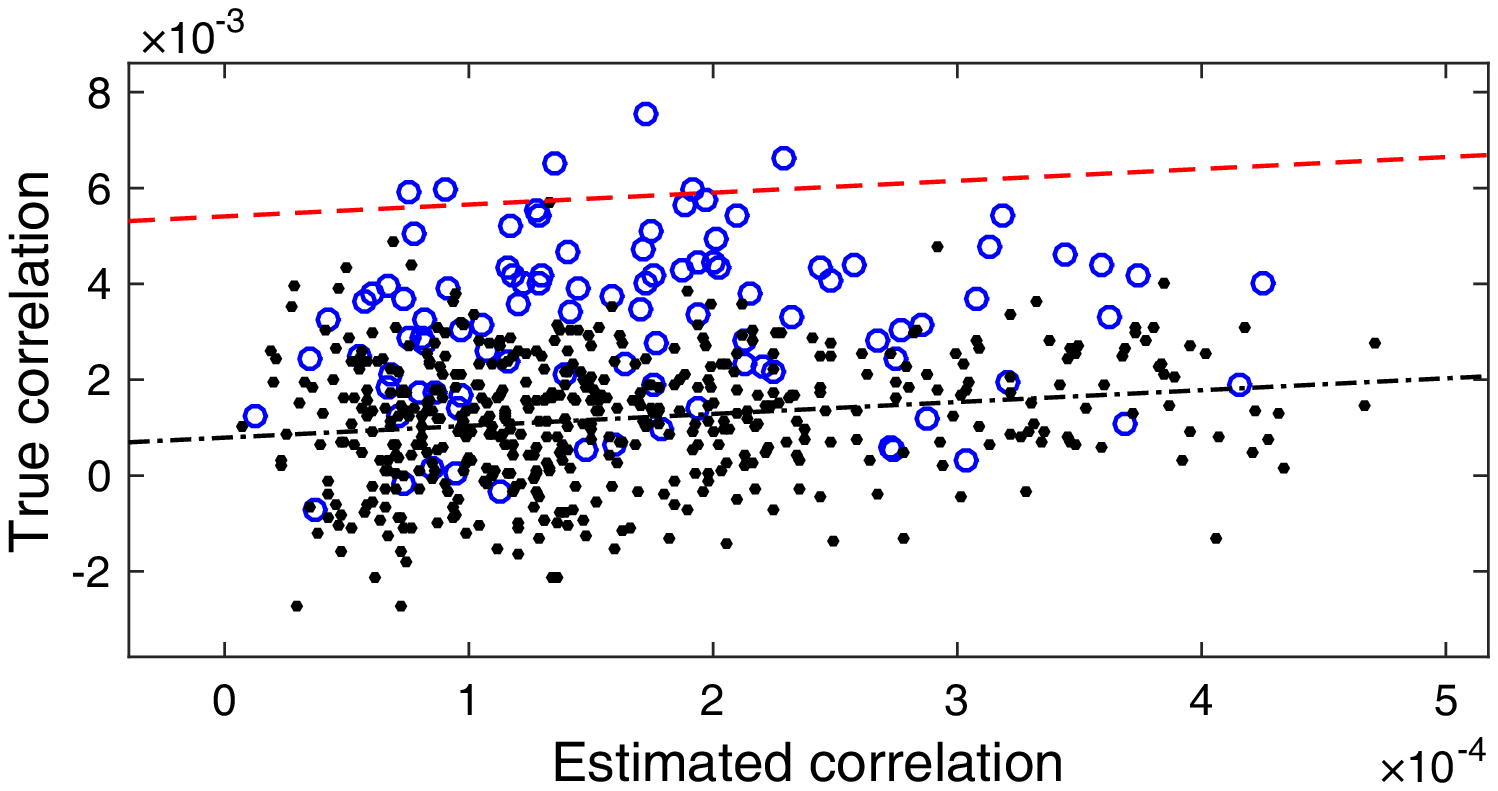}\\
  {\footnotesize (a) $l=8$}
  \end{minipage}
  \hspace{0.02\textwidth}
  \begin{minipage}[c]{0.48\textwidth}
  \centering
  \includegraphics[width=7cm]{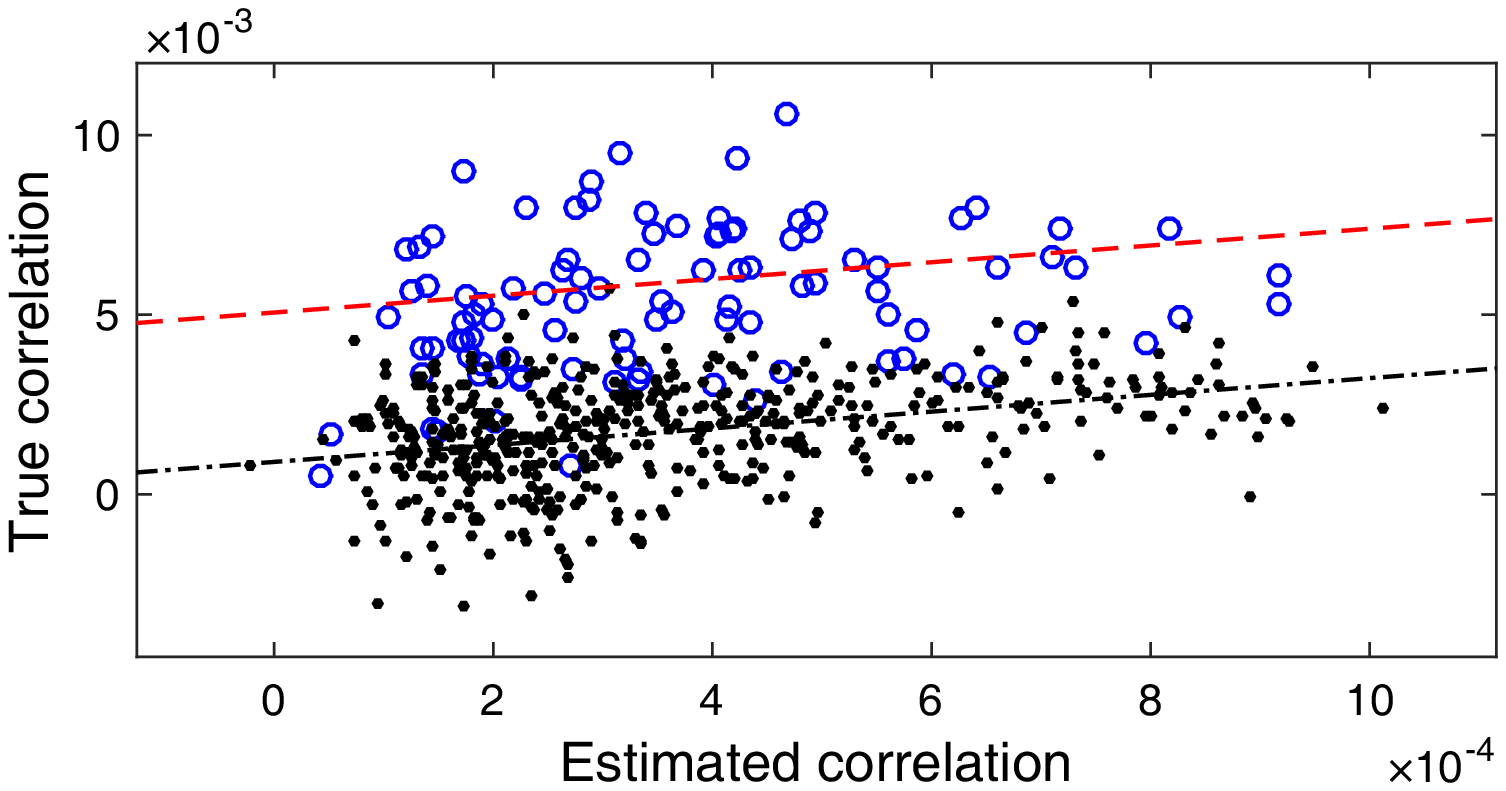}\\
  {\footnotesize (b) $l=32$}
  \end{minipage}
  \newline
  \begin{minipage}[c]{0.48\textwidth}
  \centering
  \includegraphics[width=7cm]{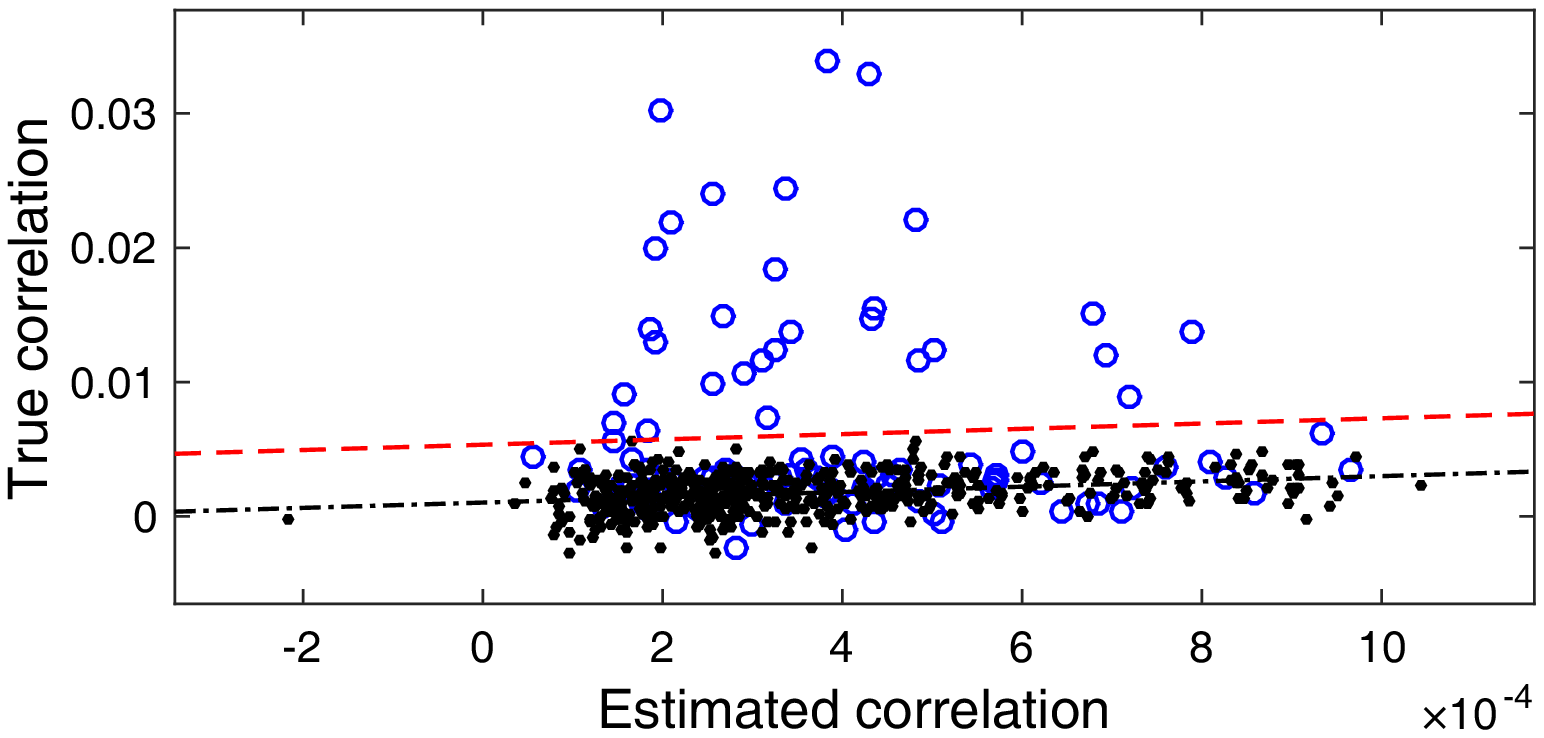}\\
  {\footnotesize (c) $l=512$}
  \end{minipage}
  \hspace{0.02\textwidth}
  \begin{minipage}[c]{0.48\textwidth}
  \centering
  \includegraphics[width=7cm]{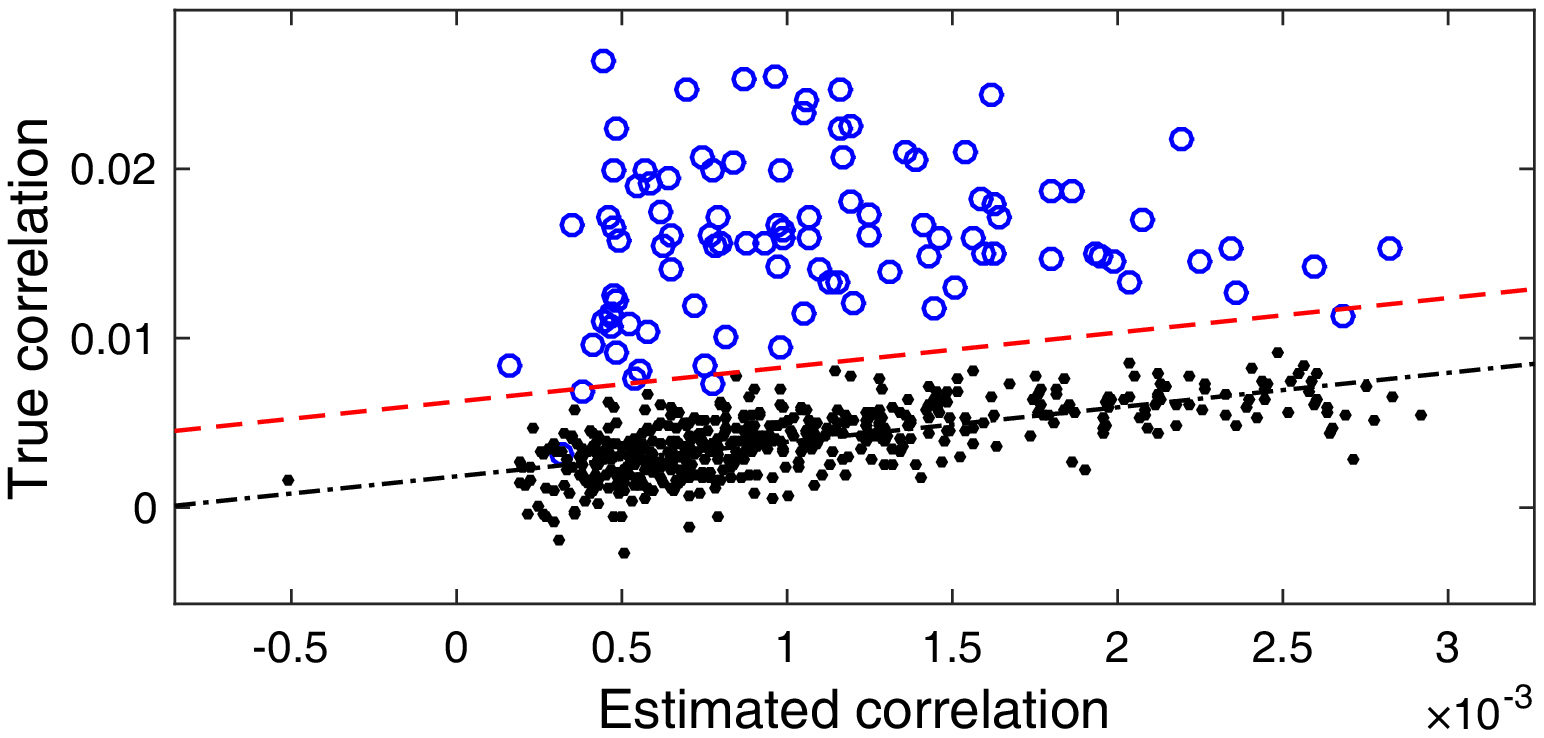}\\
  {\footnotesize (d) Conventional method ($l=1024$)}
  \end{minipage}
  \caption[True correlation $c_{{\bf I},{\bf J}^\prime}$  vs. estimated correlation $\hat c_{{\bf I},{\bf J}^\prime}$  for the forged image \#198 in $\mathcal{C}_{E1}$.]{True correlation $c_{{\bf I},{\bf J}^\prime}$  vs. estimated correlation $\hat c_{{\bf I},{\bf J}^\prime}$  for the forged image \#198 in $\mathcal{C}_{E1}$. The blue circles, black dots, and red dashed lines denote the images used by Eve, the images not used by Eve, and the detection thresholds, respectively. The dash-dotted lines are fitted lines of the black dots. Please note that the results for other images are similar based on our experiments.}\label{figure:select_l}
\end{figure*}

In order to reduce the computational complexity, all images are center cropped with the size of $1024\times1024$, which are then JPEG compressed with a high quality factor 90 just as did in \cite{Goljan2011} to avoid the inconsistence of JPEG quantization matrices. We assume that the attacker Eve steals $N$ images from $\mathcal{C}_A$ and creates some forgeries for each target image from $\mathcal{C}_{E1}$, $\mathcal{C}_{E2}$ and $\mathcal{C}_{E3}$ using the conventional fingerprint-copy attack \cite{Goljan2011}, and the proposed method with different parameters. For a fair comparison, we also adjust the $\alpha$ in Eq. (\ref{eqt:fingerprint_copy}) based on PSNR like Eq. (\ref{eqt:alpha_b}) when applying the conventional method, since Eve cannot obtain the true fingerprint $\bf K$  of camera  $\mathcal{C}_A$ in real forensic cases. As it did in \cite{Goljan2011}, the target images are slightly denoised by the wavelet filter \cite{Mihcak1999} with parameter $\sigma=1$ to suppress their own fingerprints and other possible artifacts before superimposing the fake estimated fingerprint.

\subsection{Parameter Selection}
\label{subsec:parameter}

\subsubsection{Selection of Parameter $l$}
Since the proposed method works in a block-wise manner, the parameter $l$ (\emph{i.e.} block size) would have impact on the anti-forensic performance. In order to select a proper value of $l$, we have conducted the following experiments.

By setting $N$=100, $A$=50, $r$=10, and $l\in\{8,16,32,64,128,256,512\}$, we create 300$\times$7 (with regard to different $l$) forged images from camera $\mathcal{C}_{E1}$ using the proposed method. Then we investigate the performance of camera identification and individual triangle test, respectively. For the forged images, we aim to increase the false alarm rate of camera identification ($P_{fa}$) and decrease the correct detection rate of triangle test ($P_{D}$). The experimental results are shown in Table~\ref{table:select_l}. From Table~\ref{table:select_l}, it is observed that we obtain $P_{fa}=100\%$ when $l\geq32$, meaning that the camera identification is successfully deceived. When $l$ is very small (\emph{e.g.} 8) or very large (\emph{e.g.} 512), the obtained $P_D$ of triangle test is relatively smaller than those obtained with a “middle range” $l$ (\emph{e.g.} 1024 or 256). Due to a small $l$ would lead to poor performance for deceiving the camera identification, it may be better to choose a large $l$.

However, our further analysis indicates that a too large $l$ would bring some potential risks. As illustrated in Fig.~\ref{figure:select_l}, the true correlation values (blue circles) for relatively small $l$ (\emph{i.e.} Fig.~\ref{figure:select_l}(a) and Fig.~\ref{figure:select_l}(b)) are always close to the detection threshold (red dashed line), and most of them are less than $8\times10^{-3}$. However, some of the true correlation values for a large $l$ (\emph{i.e.} Fig.~\ref{figure:select_l}(c)) are significantly larger than the threshold. Comparing Fig.~\ref{figure:select_l}(a) and Fig.~\ref{figure:select_l}(b), it is observed that many true correlation values (larger than $1\times10^{-2}$ in this experiment) obtained with $l=512$ are even larger than those obtained with the conventional method ($l=1024$). These values can be regarded as outlier data, which can be used as a strong sign for image forgery. Such a case is not shown if we just focus on the $P_D$ of triangle test, but the one who attempts to perform anti-forensics must pay attention to it in a real scenario. To avoid such traces, $l$ should not be too large. Thus, we select $l=32$ in the following experiments.

\subsubsection{Selection of Parameter $A$ and $r$}
There are two important parameters in the proposed method, \emph{i.e.}, the target PSNR $A$ for setting the parameter $\alpha _b^*$ in Eq. (\ref{eqt:alpha_b}) and the number of  randomly selected images $r$.

It is noted that the PSNR between the images before and after the fingerprint-copy attack would approximately fall in the range of [47.6 dB, 58.7 dB] (please refer to Table I in \cite{Goljan2011}). Therefore, we set $A$ as 50 dB and 55 dB to evaluate the performance of the proposed method. Since the proposed method ensures the PSNR for each image block is as close to $A$ as possible, the PSNR for the whole image is also very close to the $A$.

Another parameter need to be determined is $r$. To confuse the threshold-based correlation detector, we should guarantee that the correlation between the noise residual ${\bf W}_{{\bf J}^\prime}$ of the forged image and the true fingerprint $\bf K$ of camera $\mathcal{C}_A$ is large enough. Usually, for a given $A$ (\textit{i.e.} the strength of the estimated fingerprint superimposed into $\bf J$ is given), the more stolen images we used for estimating the fingerprint, the larger the correlation we can obtain, while the poorer performance for attacking the triangle test (refer to Table~\ref{table:pd_individual} for more details). In order to select a proper $r$, the following experiments are conducted.

\begin{figure}[t]
\centering
\includegraphics[width=9cm]{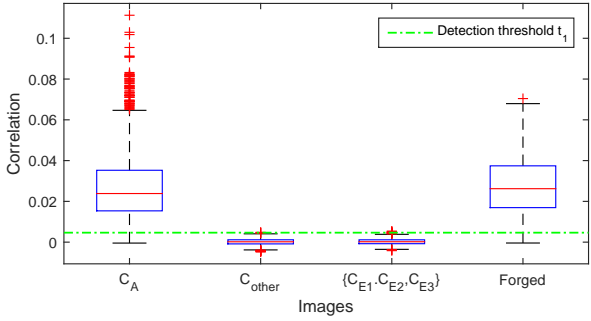} \\
\caption[Boxplots of the correlation values for images from different datasets.]{Boxplots of the correlation values for images from different datasets (the forged images are created with $N=100$, $A=50$, $r=10$). }\label{fig:correlation}
\end{figure}

Firstly, we determine the threshold $t_1$ for the threshold-based correlation detector. We collect 300 blue sky images taken by camera $\mathcal{C}_A$ and obtain a good estimation $\hat{\bf K}$ of the true fingerprint of camera $\mathcal{C}_A$. Then we use the 2,000 images from camera $\mathcal{C}_A$ as positive samples and collect 2,000 images from the Dresden Image Database (available at http://forensics.inf.tu-dresden.de/ddimgdb) taken by 10 different cameras (including 5 different camera models, denote as $\mathcal{C}_{other}$) as negative samples. We calculate the correlations of these images with $\hat{\bf K}$ according to Eq. (\ref{eqt:corr}). Finally, we set the false alarm rate $P_{fa}$ as $10^{-3}$ to obtain the threshold $t_1$ ($t_1=0.0046$ in this experiment). If the correlation of an image is larger than $t_1$, the detector would identify the image as being taken by camera $\mathcal{C}_A$. In Fig.~\ref{fig:correlation}, the boxplots of the correlations for all the images from different cameras are illustrated. It is observed that almost all images from camera $\mathcal{C}_A$ (see the leftmost boxplot) and from other cameras (\emph{i.e.} $\mathcal{C}_{other}$, $\mathcal{C}_{E1}$, $\mathcal{C}_{E2}$ and $\mathcal{C}_{E3}$, see the two boxplots in the middle) are clearly separated by the threshold $t_1$.

\begin{figure}[t]
\centering
\includegraphics[width=9cm]{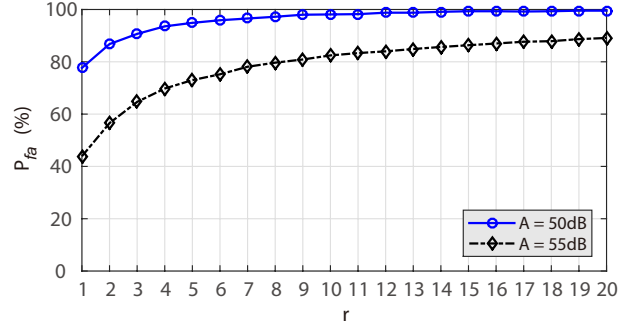} \\
\caption{$P_{fa}$ of 900 forged images for the camera identification with increasing the parameter $r$ ranging from 1 to 20.}\label{fig:r_pass}
\end{figure}

By setting $N=100$, we try to analyze the effect of the parameter $r$ on the correlation detection. Please note that similar results can be obtained for other values, such as $N=20, 50, 200$ and $300$ based on our experiments. Fig.~\ref{fig:r_pass} shows the ${P}_{fa}$ of the 900 forged images (\textit{i.e.}  the percentage of those forged images whose correlations are larger than the threshold $t_1$) with increasing the parameter $r$ from 1 to 20. It can be observed that the ${P}_{fa}$ would increase with increasing the parameter $r$. At the beginning, \emph{e.g.}, $r$ ranging from 1 to 9, the ${P}_{fa}$ increases sharply. When $r$ is larger than 15, it increases slightly, which means that there is no need to use all the stolen images for fingerprint estimation, since the more stolen images to be used for estimating the fake fingerprint, the more easily the triangle test detects the resulting forgeries. Based on the above analysis, we evaluate the proposed method with the parameter $r\in\{10,15\}$ in the following experiments.

\subsection{Experiment \#1: Confusing Camera Identification}
\label{subsec:confusion}

In this experiment, we will evaluate  the proposed method for confusing the camera identification. In this case, each of the image forgeries is tested by the correlation detector, and a higher value of ${P}_{fa}$ means the better anti-forensic performance.

\begin{table*}[t]
\centering
\caption{$P_{fa}$ (\%) of camera identification for the forgeries from three cameras.}{
\begin{tabular}{c|c|c|ccccc} \hline
  \multicolumn{3}{c|}{$N$}&20&50&100&200&300 \\ \hline\hline
  \multirow{6}{*}{$\mathcal{C}_{E1}$}
  &\multirow{3}{*}{$A=50$}
   & Method in \cite{Goljan2011} &{\emph{100}}&{\emph{100}}&{\emph{100}}&{\emph{100}}&{\emph{100}}\\
  && Proposed ($r=15$)        &   100      &   100      &   100      &   100       & 100     \\
  && Proposed ($r=10$)        &   100      &   99.67    &   100      &   99.67     & 99.67   \\ \cline{2-8}
  &\multirow{3}{*}{$A=55$}
   & Method in \cite{Goljan2011} &{\emph{97.67}}&{\emph{99.33}}&{\emph{100}}&{\emph{100}}&{\emph{100}}\\
  && Proposed ($r=15$)        &   96.00    &   96.00    &   96.00    &   95.67     & 95.67   \\
  && Proposed ($r=10$)        &   94.00    &   93.67    &   94.33    &   93.00     & 93.00   \\ \hline\hline
  \multirow{6}{*}{$\mathcal{C}_{E2}$}
  &\multirow{3}{*}{$A=50$}
   & Method in \cite{Goljan2011} &{\emph{99.67}}&{\emph{100}}&{\emph{100}}&{\emph{100}}&{\emph{100}}\\
  && Proposed ($r=15$)        &   99.33    &   99.33    &   99.33    &   99.33     & 99.00   \\
  && Proposed ($r=10$)        &   99.00    &   99.00    &   99.00    &   98.67     & 99.00   \\ \cline{2-8}
  &\multirow{3}{*}{$A=55$}
   & Method in \cite{Goljan2011} &{\emph{92.67}}&{\emph{97.67}}&{\emph{98.67}}&{\emph{99.00}}&{\emph{100}}\\
  && Proposed ($r=15$)        &   83.00    &   84.00    &   85.33    &   82.67     & 82.33   \\
  && Proposed ($r=10$)        &   79.33    &   79.33    &   80.33    &   76.67     & 78.00   \\ \hline\hline
  \multirow{6}{*}{$\mathcal{C}_{E3}$}
  &\multirow{3}{*}{$A=50$}
   & Method in \cite{Goljan2011} &{\emph{99.67}}&{\emph{99.67}}&{\emph{99.67}}&{\emph{99.67}}&{\emph{99.67}}\\
  && Proposed ($r=15$)        &   99.00    &   98.67    &   99.33    &   99.00     & 98.67   \\
  && Proposed ($r=10$)        &   98.33    &   98.00    &   98.00    &   98.00     & 97.67   \\ \cline{2-8}
  &\multirow{3}{*}{$A=55$}
   & Method in \cite{Goljan2011} &{\emph{89.67}}&{\emph{96.00}}&{\emph{99.00}}&{\emph{99.67}}&{\emph{99.67}}\\
  && Proposed ($r=15$)        &   86.67    &   86.33    &   87.33    &   84.67     & 85.33   \\
  && Proposed ($r=10$)        &   80.67    &   82.00    &   82.33    &   81.33     & 81.00   \\ \hline
\end{tabular}}
\label{table:correlation ratio}
\end{table*}

\begin{figure*}[t]
  \centering
  \begin{minipage}[c]{0.48\textwidth}
  \centering
  \includegraphics[width=7cm]{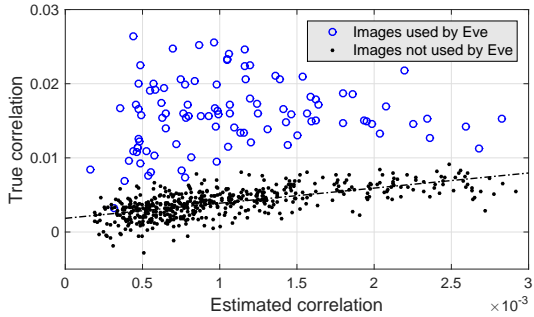}\\
  {\footnotesize (a) Conventional method}
  \end{minipage}
  \hspace{0.02\textwidth}
  \begin{minipage}[c]{0.48\textwidth}
  \centering
  \includegraphics[width=7cm]{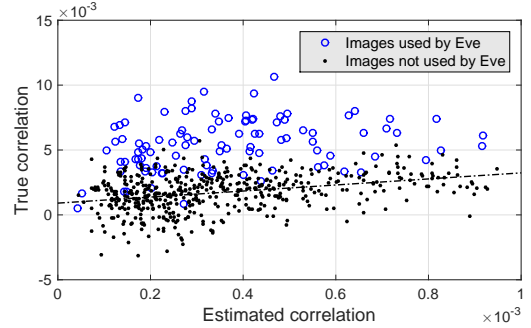}\\
  {\footnotesize (b) Proposed method}
  \end{minipage}
  \caption[Correlations $c_{{\bf I},{\bf J}^\prime}$  vs. $\hat c_{{\bf I},{\bf J}^\prime}$  for the forged image shown in Fig.~\ref{fig:forged example} using the conventional method \cite{Goljan2011} and the proposed method.]{Correlations $c_{{\bf I},{\bf J}^\prime}$  vs. $\hat c_{{\bf I},{\bf J}^\prime}$  for the forged image shown in Fig.~\ref{fig:forged example} using the conventional method \cite{Goljan2011} and the proposed method. The parameters are $N=100$, $r=10$, $A=50$ dB. The dash-dotted line is the fitted straight line of the black dots.}\label{figure:triangle test 1}
\end{figure*}

The experimental results evaluated on those forgeries from the three different cameras are shown in Table~\ref{table:correlation ratio}. It can be observed that the parameter $A$ is the most influential factor for the detection performance, since it determines the fingerprint strength. Usually, the smaller $A$ we use, the larger the fingerprint strength is, and thus the better performance for confusing the camera identification will be achieved. When $A=50$ dB, both the conventional method and the proposed method can achieve very high ${P}_{fa}$ (all larger than 97\%). As an example, we show the correlations for the forged images from $\mathcal{C}_{E1}$, $\mathcal{C}_{E2}$ and $\mathcal{C}_{E3}$ made by the proposed method with $N=100$, $A=50$, $r=10$ in the rightmost boxplot in Fig.~\ref{fig:correlation}. It is observed that the distribution of correlations for the forged images and that for the images from $\mathcal{C}_A$ are quite similar. However, the ${P}_{fa}$ will decrease when $A$ increases to 55 dB, and the decrements for the proposed method are severer comparing with the conventional method. Besides, we find that the ${P}_{fa}$ changes slightly for the parameter $N$. Based on Table~\ref{table:correlation ratio}, giving $A$ and $r$, the changes of ${P}_{fa}$ for the proposed method are less than 4\% when $N$ ranges from 20 to 300. Overall, the experimental results show that we can set $A$ as 50 dB to make almost all forged images successfully confuse the camera identification.

\subsection{Experiment \#2: Attacking Individual Triangle Test}
\label{subsec:Individual}

In this experiment, we will evaluate the proposed method for attacking individual triangle test. In this scenario, Alice try to detect the forgeries via identifying those images stolen by Eve, so a good anti-forensic method should reduce the detection rate $P_{\rm D}$ of individual test as far as possible. In our experiments, we firstly estimate an approximate fingerprint $\hat{\bf K}_A$ using $N_A=70$ blue sky images from camera $\mathcal{C}_A$. \cite{Goljan2011} has pointed out that it is not necessary for Alice to work with a better quality fingerprint than Eve, and $N_A$ is recommended from 15 to 70.  For each forged image ${\bf J}^\prime$, the individual test is performed on 600 images taken by camera $\mathcal{C}_A$. Among the 600 images, $N$ ($N\in\{20,50,100,200,300\}$ in our experiments) stolen images used by Eve are treated as candidate images, and the remaining $600-N$ images are used to estimate the pdf $f_{{\bf J}^\prime}(x)$ for ${\bf J}^\prime$ (refer to Section \ref{subsec:triangle} for the details of $f_{{\bf J}^\prime}(x)$).

\begin{table*}[!t]
\centering
\caption{The average $P_{\rm D}$ (\%) of individual triangle test for the forgeries from three cameras.}{
\begin{tabular}{c|c|c|ccccc} \hline
  \multicolumn{3}{c|}{$N$}&20&50&100&200&300 \\ \hline\hline
  \multirow{6}{*}{$\mathcal{C}_{E1}$}
  &\multirow{3}{*}{$A=50$}
   & Method in \cite{Goljan2011} &{\emph{99.30}}&{\emph{97.74}}&{\emph{92.10}}&{\emph{80.12}}&{\emph{67.48}}\\
  && Proposed ($r=15$)        &   98.07    &   89.55    &   62.97    &   26.30     & 11.65   \\
  && Proposed ($r=10$)        &   97.02    &   85.22    &   54.19    &   18.06     & 7.25    \\ \cline{2-8}
  &\multirow{3}{*}{$A=55$}
   & Method in \cite{Goljan2011} &{\emph{87.20}}&{\emph{78.72}}&{\emph{63.80}}&{\emph{41.80}}&{\emph{27.58}}\\
  && Proposed ($r=15$)        &   82.67    &   50.47    &   18.95    &   3.17      & 1.20    \\
  && Proposed ($r=10$)        &   77.33    &   39.65    &   12.03    &   1.98      & 0.88    \\ \hline \hline
  \multirow{6}{*}{$\mathcal{C}_{E2}$}
  &\multirow{3}{*}{$A=50$}
   & Method in \cite{Goljan2011} &{\emph{97.57}}&{\emph{92.86}}&{\emph{82.39}}&{\emph{64.44}}&{\emph{51.60}}\\
  && Proposed ($r=15$)        &   95.95    &   77.79    &   43.79    &   13.22     & 5.13    \\
  && Proposed ($r=10$)        &   93.65    &   70.11    &   34.38    &   8.34      & 3.07    \\ \cline{2-8}
  &\multirow{3}{*}{$A=55$}
   & Method in \cite{Goljan2011} &{\emph{75.82}}&{\emph{60.28}}&{\emph{43.75}}&{\emph{23.92}}&{\emph{14.87}}\\
  && Proposed ($r=15$)        &   67.57    &   31.38    &   8.41     &   1.51      & 0.69    \\
  && Proposed ($r=10$)        &   60.23    &   23.09    &   4.98     &   0.95      & 0.47    \\ \hline\hline
  \multirow{6}{*}{$\mathcal{C}_{E3}$}
  &\multirow{3}{*}{$A=50$}
   & Method in \cite{Goljan2011} &{\emph{97.92}}&{\emph{95.15}}&{\emph{88.58}}&{\emph{75.23}}&{\emph{63.08}}\\
  && Proposed ($r=15$)        &   96.37    &   85.73    &   59.14    &   23.22     & 9.83    \\
  && Proposed ($r=10$)        &   94.47    &   80.33    &   50.10    &   16.22     & 6.26    \\ \cline{2-8}
  &\multirow{3}{*}{$A=55$}
   & Method in \cite{Goljan2011} &{\emph{83.00}}&{\emph{75.35}}&{\emph{61.55}}&{\emph{40.68}}&{\emph{28.71}}\\
  && Proposed ($r=15$)        &   77.52    &   50.59    &   20.41    &   3.58      & 1.39    \\
  && Proposed ($r=10$)        &   72.45    &   41.28    &   13.59    &   2.06      & 0.86    \\ \hline
\end{tabular}}
\label{table:pd_individual}
\end{table*}

In Fig.~\ref{figure:triangle test 1}, we plot the $c_{{\bf I},{\bf J}^\prime}$ versus $\hat c_{{\bf I},{\bf J}^\prime}$ for two forgeries created from the image example illustrated in Fig.~\ref{fig:forged example} with the conventional method \cite{Goljan2011} and the proposed method, respectively. Here, the parameters are set as $N=100$, $A=50$ dB, and $r=10$. Compared with the Fig.~\ref{figure:triangle test 1}(a), it is observed that the separations between the images used and those not used by Eve significantly deteriorate after applying the proposed method. When $P_{\rm FA}$ (\emph{i.e.}, the false alarm rate of triangle test) is set as $10^{-3}$, the $P_{\rm D}$ for the conventional method is as high as 96\%, while the $P_{\rm D}$ for the proposed method is only 39\%.

\begin{figure*}[t]
  \centering
  \begin{minipage}[c]{0.3\textwidth}
  \centering
  \includegraphics[width=5cm]{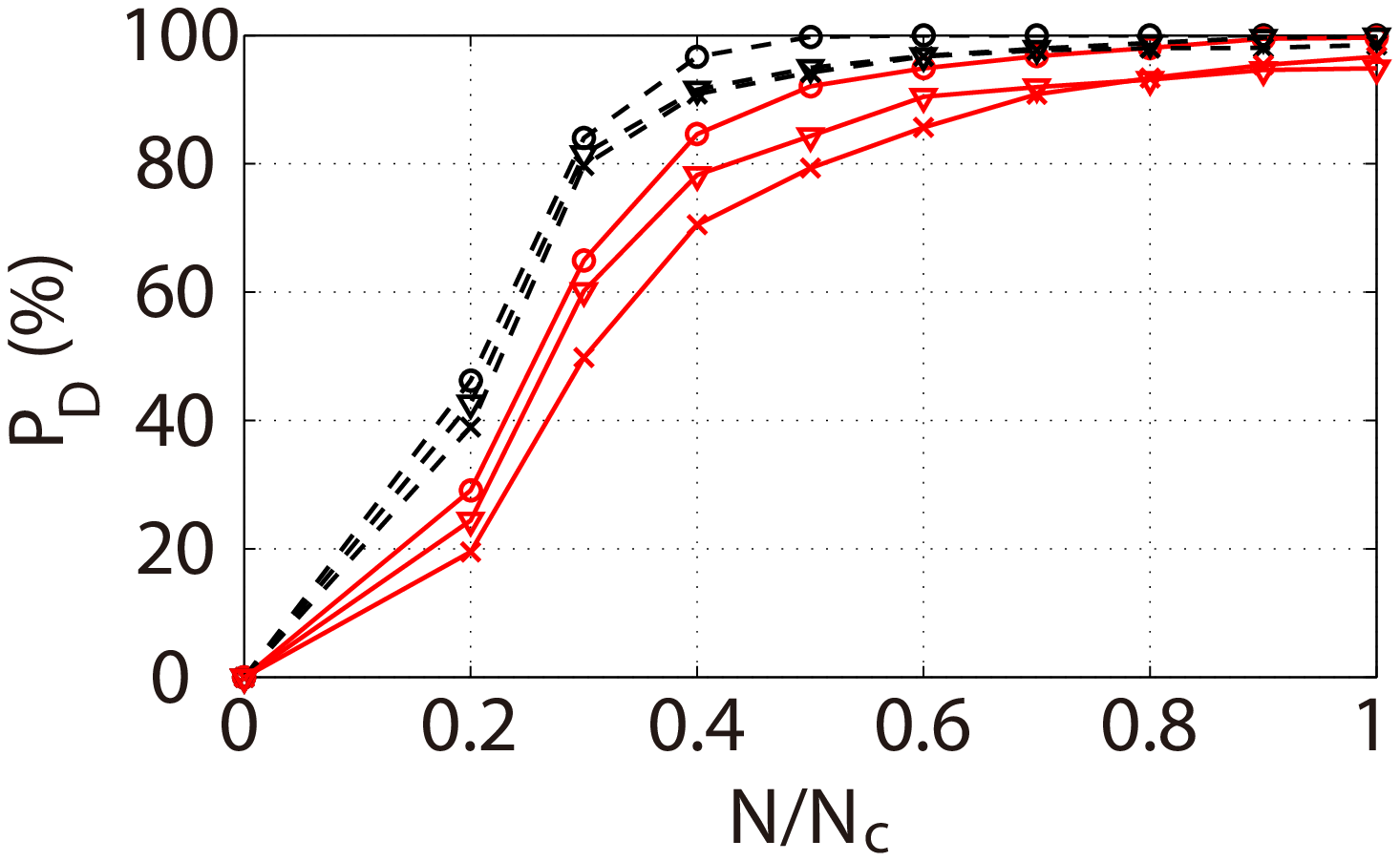}\\
  {\footnotesize (a) N=100}
  \end{minipage}
  \begin{minipage}[c]{0.3\textwidth}
  \centering\includegraphics[width=5cm]{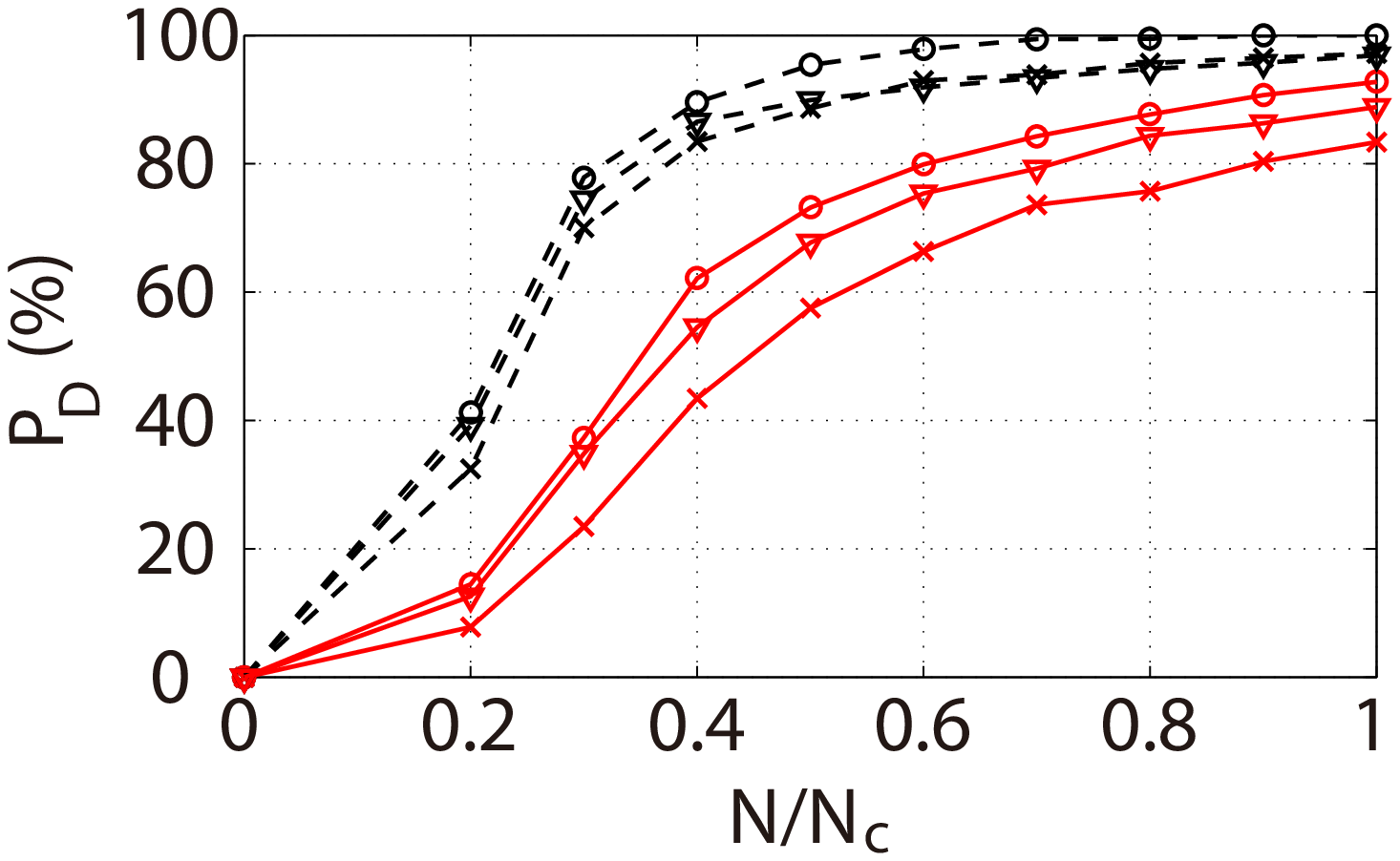}\\
  {\footnotesize (b) N=200}
  \end{minipage}
  \begin{minipage}[c]{0.3\textwidth}
  \centering
  \includegraphics[width=5cm]{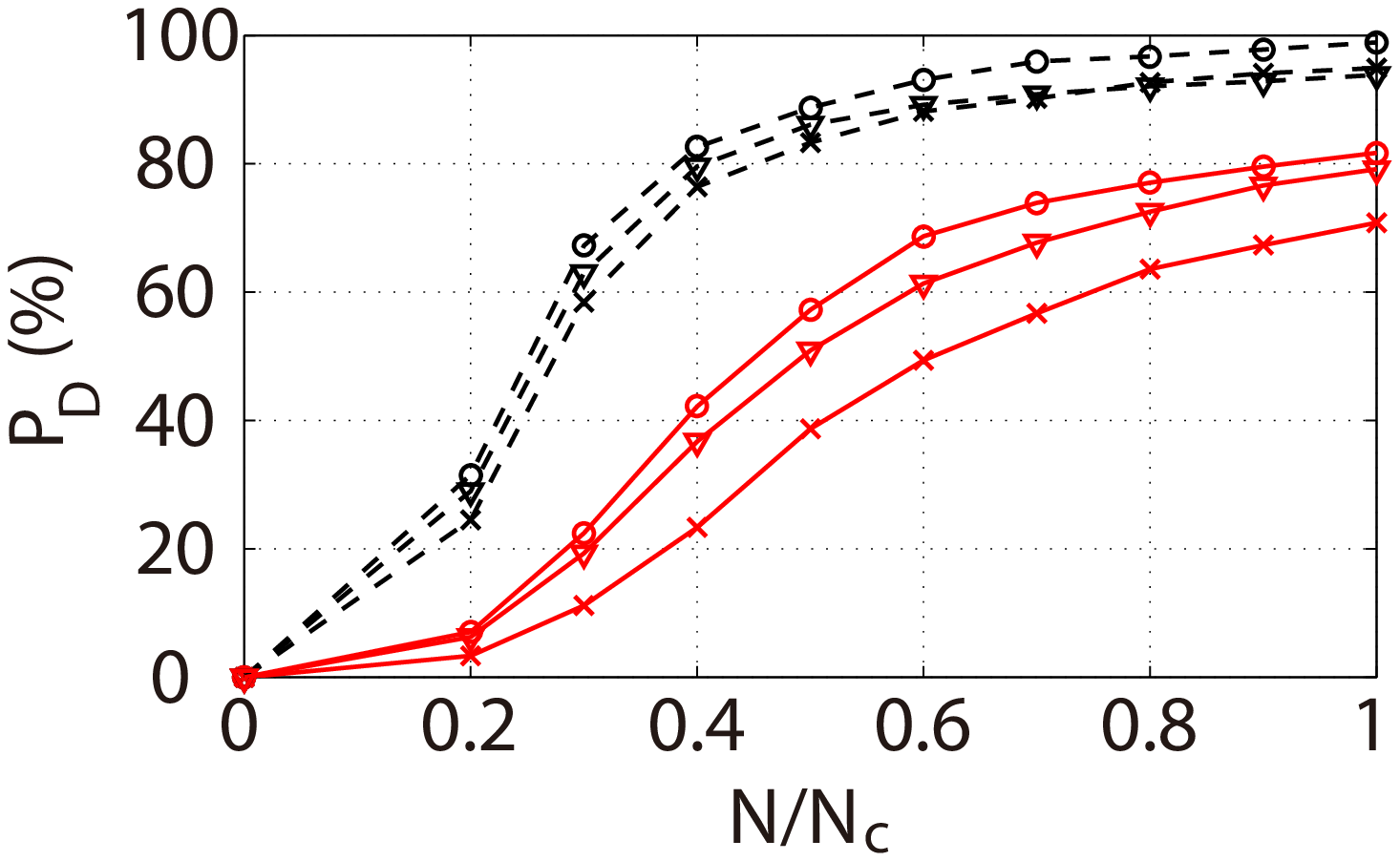}\\
  {\footnotesize (c) N=300}
  \end{minipage}
  \newline
  \includegraphics[scale=0.5]{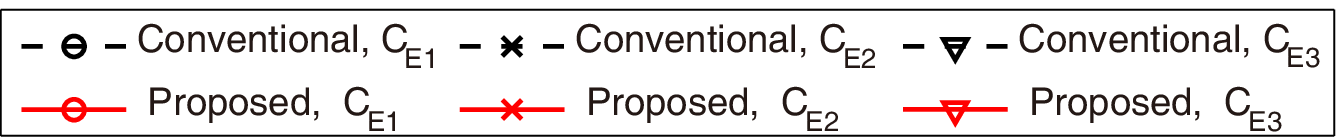}
  \caption[The average $P_{\rm D}$ of the pooled test for the conventional method \cite{Goljan2011} and the proposed method as a function of the ratio $N/N_c$ on 100 images at $P_{\rm FA}=10^{-3}$.]{The average $P_{\rm D}$ of the pooled test for the conventional method \cite{Goljan2011} and the proposed method as a function of the ratio $N/N_c$ on 100 images at $P_{\rm FA}=10^{-3}$, where $N$ is the number of images used by Eve to estimate the fake fingerprint, and $N_c$ is the number of candidate images. The parameters are $A=50$ dB and $r=10$.}\label{figure:pooled_test}
\end{figure*}

To further show the effectiveness of the proposed method, we conduct the individual test on all the forgeries made with different methods and parameters. The average $P_{\rm D}$ evaluated on 300 forged images from each of the three cameras are shown in Table~\ref{table:pd_individual}. Overall, the proposed method outperforms the conventional method significantly, especially when $N$ is large. For instance, when $N > 50$ and $A=50$ dB, the average decrements of $P_{\rm D}$ for the proposed method are over 48\% comparing with the conventional method. Besides, the larger the number of stolen images $N$ is, the poorer detection performance for the individual triangle test we achieve. When $N > 100$, $P_{\rm D}$ for the proposed method with $A=55$ dB are less than 4\%, meaning that the individual triangle test would almost become useless in these cases.

\subsection{Experiment \#3: Attacking Pooled Triangle Test}
\label{subsec:Pooled}

In this experiment, we will evaluate the proposed method for attacking the pooled triangle test. Firstly, we randomly select 100 target images from each of the camera $\mathcal{C}_{E1}$, $\mathcal{C}_{E2}$ and $\mathcal{C}_{E3}$ respectively, and create totally 1,800 forgeries using the selected images with the proposed method and the conventional method and the parameters $N \in \{100,200,300\}$, $A = 50~{\rm dB}$, and $r=10$. Please note that the two methods with these parameters can obtain similar good results (all $P_{fa}> 96\%$) on confusing camera identification according to the experiment \#1. For each forged image $\bf J^\prime$, the pdf $f_{{\bf J}^\prime}(x)$ is estimated from 500 images taken by camera $\mathcal{C}_A$ which are not used by Eve. Assume that the $N$ stolen images are a subset of the $N_c$ candidate images $\bf I$ from camera $\mathcal{C}_A$ ($N_c \geq N$). For a given ratio $N/N_c$, $k = 60$ images are randomly selected out of the $N_c$ candidate images and their statistics $c_{{\bf I},{\bf J}^\prime}-\lambda\hat{c}_{{\bf I},{\bf J}^\prime}-\eta$ are computed. If the $p-$value of these statistics is less than $P_{\rm FA}=10^{-3}$, we determine $\bf J^\prime$ is a forgery. Like what was done in \cite{Goljan2011}, we repeat the process of random selecting $k$ images and making a decision on $\bf J^\prime$ 10,000 times, and finally obtain the $P_{\rm D}$ for $\bf J^\prime$, \emph{i.e.}, the probability of that $\bf J^\prime$ is correctly detected as a forgery over the 10,000 times.

Fig.~\ref{figure:pooled_test} shows the ratio $N/N_c$ (please note that $N$ is set as 100, 200 or 300, $N_c$ would change with different ratio of $N/N_c$.) versus the average $P_{\rm D}$ on 100 images from the three cameras. It is observed that the conventional method \cite{Goljan2011} can be easily detected by the pooled test, especially when $N/N_c>0.5$ ($P_{\rm D}>88\%$ in such cases). However, the proposed method can decrease the $P_{\rm D}$ significantly. For a given $N$, here $N=100, 200, 300$, the decrements of $P_{\rm D}$ averaging over the three cameras with all ratio $N/N_c$ are 10.0\%, 22.1\%, and 29.9\%, respectively. From Fig.~\ref{figure:pooled_test}, it is also observed that the $P_{\rm D}$ will increase when the ratio $N/N_c$ ranging from 0 to 1. When $N/N_c \leq 0.4$ and $N \geq 200$, the average $P_D$ is smaller than 53.4\% for the proposed method. Even when $N/N_c=1$ (\textit{i.e.} the set of the stolen images for Eve is exactly the same as the set of candidate images used in the pool test for Alice, and this case seems unlikely to happen in practice for detectors), the average $P_{\rm D}$ for the proposed method is still less than 88.3\% when $N\geq200$, while the average $P_{\rm D}$ for the conventional method \cite{Goljan2011} is larger than 95.9\%.

\begin{table*}[t]
\centering
\caption{The average $P_{\rm D}$ (\%) of multiple forgeries triangle test for the forgeries from three cameras.}{
\begin{tabular}{c|c|c|ccccc} \hline
  \multicolumn{3}{c|}{$N$}&20&50&100&200&300 \\ \hline\hline
  \multirow{6}{*}{$\mathcal{C}_{E1}$}
  &\multirow{3}{*}{$A=50$}
   & Method in \cite{Goljan2011} &{\emph{99.97}}&{\emph{99.97}}&{\emph{100}}&{\emph{99.90}}&{\emph{99.97}}\\
  && Proposed ($r=15$)        &   99.73    &   93.61    &   76.19    &   54.98     & 42.11   \\
  && Proposed ($r=10$)        &   98.36    &   86.15    &   63.11    &   39.13     & 24.82   \\ \cline{2-8}
  &\multirow{3}{*}{$A=55$}
   & Method in \cite{Goljan2011} &{\emph{73.78}}&{\emph{71.87}}&{\emph{74.52}}&{\emph{69.97}}&{\emph{72.54}}\\
  && Proposed ($r=15$)        &   66.92    &   35.15    &   16.02    &   4.38      & 2.37    \\
  && Proposed ($r=10$)        &   52.81    &   21.91    &   6.19    &   2.64      & 1.64    \\ \hline \hline
  \multirow{6}{*}{$\mathcal{C}_{E2}$}
  &\multirow{3}{*}{$A=50$}
   & Method in \cite{Goljan2011} &{\emph{99.50}}&{\emph{99.13}}&{\emph{98.90}}&{\emph{98.39}}&{\emph{98.09}}\\
  && Proposed ($r=15$)        &   99.46    &   91.10    &   74.05    &   60.10     & 53.01   \\
  && Proposed ($r=10$)        &   97.16    &   82.11    &   63.61    &   51.07     & 47.53   \\ \cline{2-8}
  &\multirow{3}{*}{$A=55$}
   & Method in \cite{Goljan2011} &{\emph{72.74}}&{\emph{72.41}}&{\emph{72.51}}&{\emph{71.47}}&{\emph{71.74}}\\
  && Proposed ($r=15$)        &   68.43    &   51.40    &   45.52    &   44.45     & 44.21   \\
  && Proposed ($r=10$)        &   60.00    &   46.92    &   44.35    &   42.51     & 44.05   \\ \hline \hline
  \multirow{6}{*}{$\mathcal{C}_{E3}$}
  &\multirow{3}{*}{$A=50$}
   & Method in \cite{Goljan2011} &{\emph{99.30}}&{\emph{99.23}}&{\emph{99.20}}&{\emph{99.00}}&{\emph{98.70}}\\
  && Proposed ($r=15$)        &   99.23    &   95.59    &   90.57    &   81.00     & 72.91   \\
  && Proposed ($r=10$)        &   98.13    &   92.31    &   84.58    &   74.11     & 67.53   \\ \cline{2-8}
  &\multirow{3}{*}{$A=55$}
   & Method in \cite{Goljan2011} &{\emph{89.36}}&{\emph{89.60}}&{\emph{89.97}}&{\emph{89.00}}&{\emph{87.32}}\\
  && Proposed ($r=15$)        &   88.33    &   74.88    &   65.72    &   60.77     & 59.73   \\
  && Proposed ($r=10$)        &   83.68    &   70.30    &   62.58    &   58.70     & 60.03   \\ \hline
\end{tabular}}
\label{table:pd_multi}
\end{table*}

\subsection{Experiment \#4: Attacking Multiple Forgeries Triangle Test}
\label{subsec:mutiple test}

In this experiment, we will evaluate the proposed method for attacking the multiple forgeries triangle test. The number of the images stolen by Eve is set as $N\in\{20,50,100,200,300\}$, where the $N$ stolen images are belong to 600 images from camera $\mathcal{C}_A$. For each forgery $\bf{J'}$, the pdf $f_{{\bf J}^\prime}(x)$ is estimated by $600-N$ images not used by Eve just as we did in experiment \#2, and the candidate images in this case are the other 299 (=300-1) forgeries from the same camera generated with the same method and parameters. Thus the $P_{\rm D}$ of $\bf{J'}$ means the probability of that the 299 forgeries are correctly detected by the multiple forgeries test.
For each method with some given parameters, we perform multiple forgeries test on 100 forgeries from camera $\mathcal{C}_{E1}$, $\mathcal{C}_{E2}$ and $\mathcal{C}_{E3}$ respectively. The average $P_{\rm D}$ are listed in the Table~\ref{table:pd_multi}.

From Table~\ref{table:pd_multi}, it is observed that the $P_{\rm D}$ for the conventional method with $A=50$ dB are larger than 98\% even when $N=300$. Overall, the proposed method can effectively degrade the $P_{\rm D}$, especially when $N$ is large. Taking $A=50$, $N=300$ and $r=10$ as an example, the average decrement for the proposed method is 52.3\% comparing to the conventional method.
It is also observed that the proposed method performs better on camera $\mathcal{C}_{E1}$ than camera $\mathcal{C}_{E2}$ and $\mathcal{C}_{E3}$. For example, when $N=100$ and $r=10$, the $P_{\rm D}$ of the proposed method for camera $\mathcal{C}_{E1}$ is 6.19\%, while the $P_{\rm D}$ for camera $\mathcal{C}_{E2}$ and $\mathcal{C}_{E3}$ are 44.35\% and 62.58\%, respectively. The possible reason may be the effects of sensor noises and/or compression artifacts introduced by the digital cameras. Since camera $\mathcal{C}_{E1}$ and $\mathcal{C}_A$ are with the same model, most hardware and software employed within the two cameras may be very similar or exactly the same. Therefore, it is expected that the non-PRNU components of their resulting images are relatively similar compared with those images from different brands and models of camera $\mathcal{C}_{E2}$ and $\mathcal{C}_{E3}$, thus the forgeries from camera $\mathcal{C}_{E2}$ and $\mathcal{C}_{E3}$ are easier to be detected. In our future research, we will further analyze the key factors inside different cameras that influence the detection performance of the multiple forgeries triangle test.

\section{Concluding Remarks}
\label{Sec:Conclusion}

In this paper, we first analyze the limitations of the conventional fingerprint-copy attack, and then propose an improved fingerprint-copy attack scheme via estimating the fake fingerprint from a randomly selected subset of the stolen images and superimposing it into the target image in a dispersive manner. The proposed scheme achieves a good tradeoff between the two requirements as described in Section \ref{Sec:Proposed_Method}. The extensive experimental results evaluated on 2,900 images from 4 different cameras have shown that the proposed method can successfully confuse the camera identification (refer to Experiment \#1) and attack the triangle test in three different forensic scenarios (refer to Experiments \#2, \#3 and \#4 respectively). It significantly outperforms the conventional fingerprint-copy attack \cite{Goljan2011}, especially when the number of stolen images $N$ is large enough, e.g. $N>100$.

Though the proposed method is effective, it still has some limitations. Firstly, like most current anti-forensics works aiming at some certain forensic methods (such as \cite{Stamm2011}, \cite{Gloe2007}, and \cite{Caldelli2011}), the proposed method is designed against one of the most popular camera identification methods \cite{Chen2008} and the triangle test \cite{Goljan2011}. Thus it is difficult to attack other camera identification methods that are not based on PRNU. Secondly, the local textural information within the image has not been fully considered in the proposed scheme. It is promising that one can obtain better anti-forensic results by setting different fingerprint strengths for different image blocks according to the textural complexity. In the future, we will try to develop an adaptive way for adjusting the fingerprint strength to further improve the proposed scheme.


\end{document}